\documentclass[hyper]{JHEP} % 10pt is ignored!

\usepackage{epsfig}
\newcommand\fverb{\setbox\pippobox=\hbox\bgroup\verb}

\newcommand\fverbdo{\egroup\medskip\noindent%

            \fbox{\unhbox\pippobox}\ }

\newcommand\fverbit{\egroup\item[\fbox{\unhbox\pippobox}]}

\newbox\pippobox

\title{Note About Null Dimensional Reduction of M5-Brane}
\author{J. Kluso\v{n}\\
Department of
Theoretical Physics and Astrophysics\\
Faculty of Science, Masaryk University\\
Kotl\'{a}\v{r}sk\'{a} 2, 611 37, Brno\\
Czech Republic\\
E-mail: \email{klu@physics.muni.cz}} \preprint{}

 \abstract{In this short note we study null dimensional reduction 
 of M5-brane covariant action. We analyse longitudinal dimensional reduction that
leads to non-relativistic D4-brane and transverse reduction that leads 
to NS5-brane in non-relativistic string theory.}
\def\tH{\tilde{H}}

\def\hv{\hat{v}}

\def\hH{\hat{H}}
\def\hV{\hat{V}}

\def\hE{\hat{E}}

\def\hE{\hat{E}}
\def\tF{\tilde{F}}

\def\barh{\bar{h}}

\def\be{\begin{equation}}

\def\ee{\end{equation}}

\def\bea{\begin{eqnarray}}
\def\bh{\bar{h}}

\def\eea{\end{eqnarray}}

\def\mH{\mathcal{H}}

\def\tV{\tilde{V}}

\def\bM{\mathbf{M}}

\newcommand{\hg}{\hat{g}}

\newcommand{\mL}{\mathcal{L}}

\begin{document}
%%%%%%%%%%%%%%%%%%%%%
%%%%Introduction %%%%%%%%%
%%%%%%%%%%%%%%%%%%%%
\section{Introduction and Summary}
It has been  known from the late nineties that  string theories have
origin in eleven dimensional theory- M-theory \cite{Witten:1995ex}.
Since all string theories can be considered as limits of M-theory it is clear that
M-theory has to contain extended objects that are parent objects of fundamental
strings, D-branes and NS5-branes \cite{Townsend:1995af}, for review see for example \cite{Townsend:1996xj}. Fundamental object of M-theory is
M2-brane as string is fundamental for string theory since M2-brane electrically couples to three form $C^{(3)}$ in the same way as string couples to two form $B^{(2)}$. On the other hand more  interesting object in M-theory is M5-brane \cite{Pasti:1997gx,Bandos:1997gm,Bandos:1997ui,Bandos:2000az,Aganagic:1997zk,
Aganagic:1997zq,Witten:1996hc}. This is six dimensional object with an exceptional property that on its world-volume self-dual three form field propagates. This 
fact makes the analysis of this theory rather complicated. On the other hand covariant action for M5-brane was proposed and extensively studied in \cite{Pasti:1997gx,Bandos:1997gm,Bandos:1997ui}. In particular, it was shown in \cite{Pasti:1997gx} that this M5-brane dimensionally
reduces into D4-brane under double dimensional reduction  and it was also shown in 
\cite{Berman:1998va} that D3-brane action arises through double dimensional reduction of M5-brane action on two torus despite of the fact that the analysis is rather complicated \cite{Berman:1998va}. 

It would be certainly interesting to check properties of covariant M5-brane action in 
another situation. Such an interesting example could be null dimensional reduction of M-theory. It is well known that null dimensional reduction is rather subtle at the level of gravity action \cite{Julia:1994bs,Bergshoeff:2017dqq} however it can be performed at the level of equations of motion. Further, it is also well known that null dimensional reduction of gravity  leads to torsional Newton-Cartan gravity 
\cite{Bergshoeff:2017dqq,Hansen:2020pqs} in lower dimensional space-time. Clearly we can deduce that null dimensional reduction of M-theory would lead to torsional Newton-Cartan string theory even if the precise study of this problem has not been performed yet \footnote{See also 
	\cite{Lambert:2020scy}. For another study of non-relativistic limits of M-theory, see \cite{Blair:2021ycc,Kluson:2019uza}.}.  On the other hand we can ask the question how extended
objects in M-theory map under null dimensional reductions. In our previous paper 
\cite{Kluson:2021pux} we studied null dimensional reduction of M2-brane. We showed that when we have M2-brane wrapped null direction we get fundamental string action in torsional Newton-Cartan string theory. Further, when we consider M2-brane transverse to this null direction we get D2-brane action in the same background. Then we can ask the question how this procedure can be applied in case of M5-brane. Exactly the goal of this paper is to study this problem. To begin with we firstly review dimensional reduction along space-like circle that should lead to D4-brane in Type IIA theory. This procedure was firstly discussed in \cite{Pasti:1997gx} but we analyse this problem in more details and in slightly different way. First of all we perform spatial dimensional reduction of M5-brane that leads to five dimensional action with propagating two form on its world-volume. Since two form in five dimensions is dual to one form it is natural to perform duality transformations following \cite{Aganagic:1997zk,Tseytlin:1996it}. We show that dual action for dimensional reduced M5-brane action is D4-brane action in Type IIA theory. We would like to stress that this analysis could be considered as dual to the analysis performed in \cite{Aganagic:1997zk} where the starting point was D4-brane and then it was shown that dual action corresponds to M5-brane action dimensionally reduced along spatial dimension.

We extend this idea to the case of null dimensional reduction of M5-brane. We start with the discussion of the background with null isometry following
\cite{Julia:1994bs,Bergshoeff:2017dqq}. Then we insert this background metric into covariant M5-brane action and we obtain its dimensionally reduced form. Again since this action still depends on two form field it is natural to  perform its duality transformation. Now we should be more careful than in case
of relativistic M5-brane since the induced metric corresponds to Newton-Cartan metric
where there is no Lorentz invariance in the target space-time. For that reason we
consider more general situation with diagonal components of spatial metric $h^{\alpha\beta}$, non-zero component of time one form $\tau_\alpha$ together with non-zero components of $m_\alpha$ where meaning of these symbols will be given below. 
Then we will be able to perform duality transformation as in case of the spatial dimensional reduction and we argue that resulting action corresponds to D4-brane in torsional Newton-Cartan background. This is main result of this paper that together 
with the paper \cite{Kluson:2021pux} show that extended objects in M-theory map under null dimensional reductions to fundamental strings an D-branes in torsional Newton-Cartan string theory. It is very interesting that these actions have functionally the same form as their relativistic counterparts since they depend on the pull-back of the Newton-Cartan volume form \cite{Hansen:2020pqs}. This is also especially interesting when we compare these actions with the actions of non-relativistic strings and D-branes that are derived either by specific limiting procedure 
\cite{Bergshoeff:2019pij,Bergshoeff:2018yvt,Andringa:2012uz,Bergshoeff:2021bmc},
see also \cite{Kluson:2020kyp,Kluson:2020rij,Kluson:2018grx}, or by null reduction of relativistic string 
\cite{Harmark:2019upf,Harmark:2018cdl,Harmark:2017rpg,Kluson:2020aoq,Kluson:2019xuo,
	Kluson:2018egd}. It would be certainly very interesting to understand relations between
these string theories in more details. 

This paper is organized as follows. In the next section (\ref{second}) we review main properties of M5-brane action and perform its dimensional reduction along spatial dimension. In section (\ref{third}) we perform  dimensional reduction when M5-brane wraps null direction. Finally in section (\ref{fourth}) we briefly mention dimensional reduction of M5-brane when it is transverse to null direction.  The resulting action should corresponds to NS5-brane in torsional Newton-Cartan string theory.

\section{Covariant M5-Brane Action and Spatial Dimensional Reduction}\label{second}
In this section we introduce basic notation and study spatial dimensional reduction 
of M5-brane action. Let us  consider covariant form of M5-brane action that was introduced in 
\cite{Pasti:1997gx,Bandos:1997gm,Bandos:1997ui,Bandos:2000az}, see also \cite{Aganagic:1997zk,Aganagic:1997zq}.
Explicitly, we consider M5-brane action in the form
\footnote{We work with the metric of the signature 
$(-,+,\dots,+)$.}
\begin{equation}\label{actM5}
S=-T_{M5}\int d^6\xi \left[\sqrt{-\det \left(\hg_{mn}+i\frac{\hH^*_{mn}}{\sqrt{\widehat{\partial a\partial a}}}\right)}
-\frac{\sqrt{-\hg}}{4\widehat{\partial a\partial a}}
\hH^{*mn}H_{mnr}\partial^ra\right] \ , 
\end{equation}
where
\begin{equation}
\widehat{\partial a\partial a}=\partial_m a \hg^{mn} \
\partial_n a \ , 
\end{equation}
 and 
where
\begin{equation}
m,n=0,1,\dots,5 \ , 
\end{equation}
are vector indices of $d=6$ world-volume coordinates $\xi^m$,
\begin{equation}
M,N=0,1,\dots,10
\end{equation}
are vector indices of $D=11$ target space-time coordinates
$X^M$ 
\begin{equation}
\hg_{mn}=\partial_m X^M\hg_{MN}\partial_n X^N
\end{equation}
is the world-volume metric that is induced by embedding the five-brane
into $D=11$ background with metric $\hg_{MN}$.
Further, we have
\begin{equation}
\hH^{*}_{mn}=\frac{1}{6\sqrt{-\hg}}\hg_{mr}\hg_{ns}
	\epsilon^{rstuvw}H_{tuv}\partial_w a\ , 
\end{equation}	
where $H_{mnl}$ is the field strength for the word-volume antisymmetric tensor
$B_{mn}$
\begin{equation}
 H_{mnl}=\partial_l B_{mn}+\partial_m B_{nl}+
\partial_n B_{lm}\ . 
\end{equation}
Finally $a$ is world-volume scalar field that ensures the covariance
of the model. 

The M5-brane action is invariant under following gauge symmetries
\cite{Pasti:1997gx}. 
The first one is usual gauge symmetry
\begin{equation}
\delta a=0 \ , \quad \delta B_{mn}=\partial_m\phi_n-\partial_n\phi_m \ .
\end{equation}
There is another non-trivial gauge symmetry in the form 
\cite{Pasti:1997gx,Bandos:1997gm,Bandos:1997ui}
\begin{equation}\label{delta1}
\delta B_{mn}=\phi_m\partial_n a-\partial_m a\phi_n \ , \quad 
\delta a=0 \ . 
\end{equation}
Finally  the action is invariant under transformation \cite{Pasti:1997gx,Bandos:1997gm,Bandos:1997ui}
\begin{equation}\label{delta2}
\delta a=\phi \ , \quad 
\delta B_{mn}=\frac{\delta a}{\sqrt{\hat{\partial a\partial a}}}
\left[\hat{H}^*_{\ mn}-H_{mnp}\hg^{ps}
\frac{\partial_s a}{\sqrt{\widehat{\partial a\partial a}}}\right] \ , 
\end{equation}
where
\begin{eqnarray}
\hat{\mH}^*_{mn}=-\frac{2}{\sqrt{-\hg}}
\frac{\delta \mL_{DBI}}{\delta \hat{H}^{*mn}} \ , \quad  
\mL_{DBI}=\sqrt{-\det \left(\hg_{mn}+i\frac{\hat{H}^*_{mn}}{\sqrt{\widehat{\partial a\partial a}}}\right)} \ . 
\end{eqnarray}
The symmetries (\ref{delta1}) and (\ref{delta2}) are fundamental properties
of M5-brane action (\ref{actM5}). They ensure that $B_{mn}$ equations of motion 
reduce to self-duality conditions, see \cite{Pasti:1997gx,Bandos:1997gm,Bandos:1997ui}
for more details. 
As was shown in \cite{Pasti:1997gx}
the equation of motion for $a(\xi)$ is consequence of the equations of motion 
for $B_{mn}$ and hence it is not a new solution. As a result $a$ can be eliminated
by fixing the gauge transformations 
(\ref{delta1}). Following \cite{Pasti:1997gx} we fix $a$ by imposing 
condition
\begin{equation}\label{parta}
\partial_m a=\delta_m^5 \ .
\end{equation}
Then using (\ref{delta1})  
%Then using transformation
%\begin{equation}
%\delta B_{mn}=\frac{1}{2}(\partial_m a \phi_n-
%\partial_n a\phi_m)
%\end{equation}
we get
%\begin{equation}
%\delta B_{mn}=\frac{1}{2}(\delta_m^5\phi_n-
%\delta_n^5 \phi_m)
%\end{equation}
%that implies that for $m=5$ we get that $m\neq n$ and hence
\begin{equation}
\delta B_{5m}=\frac{1}{2}\phi_m
\end{equation}
and hence we can fix components of $B_{5m}$ to be equal to zero
\begin{equation}\label{B5m}
B_{5m}=0 \ . 
\end{equation}
We impose this condition in case of spatial and null dimensional reduction of
M5-brane.

Before we proceed to the case of null dimensional reduction we 
review spatial dimensional reduction of M5-brane.  To do this we consider
eleven dimensional line element in the form 
\begin{equation}\label{lineelement}
ds^2=e^{-\frac{2}{3}\phi}g_{\mu\nu}dx^\mu dx^\nu+
e^{\frac{4}{3}\phi}(dy-C_\mu dx^\mu)^2 \ ,
\end{equation}
where $\mu,\nu=0,1,\dots,9$. 
We impose spatial dimensional reduction when we presume that $y$ coincides
with $\xi^5$ and when all world-volume fields do not depend on $\xi^5$ except $a$ that obeys the condition (\ref{parta}). Let us denote remaining world-volume coordinates as $\xi^\alpha,\alpha,\beta,\gamma,...=0,1,\dots,4$. Then using (\ref{B5m}) and also the fact that all world-volume fields do not depend on $\xi^5$ we find that there are  non-zero components of $H_{\alpha\beta\gamma}$ only.

Then with the help of (\ref{lineelement}) we find that components of the induced metric are  equal to
\begin{equation}\label{indumet}
\hg_{\alpha\beta}=\partial_\alpha x^\mu \hg_{\mu\nu}\partial_\beta x^\nu=
e^{-\frac{2}{3}\phi}g_{\alpha\beta}+e^{\frac{4}{3}\phi}C_\alpha C_\beta \ , \quad 
\hg_{\alpha y}=-e^{\frac{4}{3}\phi}C_\alpha \ , \quad 
\hg_{yy}=e^{\frac{4}{3}\phi} \
\end{equation}
with corresponding components of the inverse metric  $\hg^{mn}$
\begin{equation}
\hg^{\alpha\beta}=e^{\frac{2}{3}\phi}g^{\alpha\beta} \ , \quad \hg^{y\alpha}
=e^{\frac{2}{3}\phi}C_\gamma g^{\gamma\beta} \ , 
\quad 
\hg^{yy}=e^{-\frac{4}{3}\phi}+e^{\frac{2}{3}\phi}
C_\alpha g^{\alpha\beta}C_\beta \ , 
\end{equation}
where $g_{\alpha\beta}$ is induced five dimensional matrix and hence
$g^{\alpha\beta}$ is its inverse metric that obeys the equation
\begin{equation}
g_{\alpha\beta}g^{\beta\gamma}=\delta_\alpha^\gamma \ . 
\end{equation}
Further, using (\ref{indumet}) we obtain
\begin{equation}
\det \hg_{mn}=e^{-2\phi}\det g_{\alpha\beta} \ . 
\end{equation}
Let us now turn our attention to $\hH_{mn}^*$ that using
(\ref{parta}) has the form
%
%
%with corresponding components of inverse world-volume metric
%\begin{eqnarray}
%\hg^{55}=\frac{\det(e^{-\frac{2}{3}\phi}g_{\alpha\beta}+e^{\frac{4}{3}\phi}C_\alpha C_\beta ) }{\det \hg_{mn}}=
%\frac{e^{-\frac{10}{3}\phi}\det g_{\alpha\beta}(1+e^{2\phi}C_\alpha g^{\alpha\beta}C_\beta)}{e^{-2\phi}\det g_{\alpha\beta}}=
%\nonumber \\
%=e^{-\frac{4}{3}\phi}(1+e^{2\phi}C_\alpha C^\alpha) \ . \nonumber \\
%\end{eqnarray}
\begin{eqnarray}
\hH^*_{mn}=\frac{1}{6\sqrt{-\hg}}
\hg_{mr}\hg_{ns}\epsilon^{rstuv5}H_{tuv}  \ . 
\end{eqnarray}
Now since $\epsilon^{mnrstu}$ is totally antisymmetric symbol we find that all non-zero components have to the form $\epsilon^{\alpha\beta\gamma\delta\omega 5}\equiv \epsilon^{\alpha\beta\gamma\delta\omega}$ so that
\begin{equation}
\hH^*_{mn}=\frac{e^{-2\phi}}{6\sqrt{-g}}\hg_{m\alpha}
\hg_{n\beta}\epsilon^{\alpha\beta\gamma\delta\omega}H_{\gamma\delta\omega} \ . 
\end{equation}
%so that we have
%\begin{eqnarray}
%\hH_{\alpha\beta}=\frac{e^{-\phi/3}}{6\sqrt{-g}}
%(g_{\alpha\alpha'}+e^{2\phi}C_\alpha C_{\alpha'}
%(g_{\beta\beta'}+e^{2\phi}C_\beta C_{\beta'})
%\epsilon^{\alpha'\beta'\gamma\delta\omega}H_{\gamma\delta\omega
%}
%\nonumber\\
%\end{eqnarray}
To proceed further we use the fact that we can write 
\begin{eqnarray}
& &\sqrt{-\det \left(\hg_{mn}+i\frac{\hH^*_{mn}}{\sqrt{\widehat{\partial a
	\partial a}}}\right)}=
\nonumber \\
& &=\sqrt{-\det\hg_{mk}}\sqrt{-\det\left(\hg^{kl}+i\frac{\hg^{kr}\hH_{rs}^*\hg^{sl}}{\sqrt{\widehat{\partial a\partial a}}}\right)}\sqrt{-\det \hg_{ln}}\  , 
\nonumber \\
\end{eqnarray}
where
\begin{equation}
\hg^{kl}+
i\frac{\hg^{kr}\hH_{rs}^*\hg^{sl}}{\sqrt{\widehat{\partial a\partial a}}}
=\hg^{kl}+
i\frac{1}{6\sqrt{-\hg}\sqrt{\widehat{\partial a\partial a}}}
\epsilon^{kltuvw}H_{tuv}\partial_w a \ . 
\end{equation}
Now in the gauge $\partial_m a=\delta_m^5$ we obtain  
%Note that the induced metric has the form 
%\begin{equation}
%\hg_{\alpha\beta}=\partial_\alpha x^\mu \hg_{\mu\nu}\partial_\beta x^\nu=
%e^{-\frac{2}{3}\phi}g_{\alpha\beta}+e^{\frac{4}{3}\phi}C_\alpha C_\beta \ , 
%\hg_{\alpha y}=-e^{\frac{4}{3}\phi}C_\alpha \ , 
%\hg_{yy}=e^{\frac{4}{3}\phi} 
%\end{equation}
\begin{eqnarray}
& &\det(
\hg^{kl}+
i\frac{1}{6\sqrt{-\hg}\sqrt{\widehat{\partial a\partial a}}}
\epsilon^{kltuvw}H_{tuv}\partial_w a)=\nonumber \\
& &\left|\begin{array}{cc}
e^{\frac{2}{3}\phi}g^{\alpha\beta}+
i\frac{1}{6\sqrt{-\hg}\sqrt{\widehat{\partial a\partial a}}}
\epsilon^{\alpha\beta \gamma\delta\omega}H_{\gamma\delta\omega} &
g^{\alpha\gamma}C_\gamma e^{\frac{2}{3}\phi} \\
C_\gamma g^{\gamma\beta}e^{\frac{2}{3}\phi} &
e^{-\frac{4}{3}\phi}+e^{\frac{2}{3}\phi}
C_\alpha g^{\alpha\beta}C_\beta \\ \end{array}\right|=\nonumber \\
& &=\det 
(e^{\frac{2}{3}\phi}g^{\alpha\beta}-
\frac{e^{\frac{4}{3}\phi}g^{\alpha\gamma}C_\gamma C_\delta g^{\delta\beta}}{
e^{-\frac{4}{3}\phi}+e^{\frac{2}{3}\phi}C_\alpha g^{\alpha\beta}C_\beta}+
i\frac{1}{6e^{-\frac{5}{3}\phi}\sqrt{-\det g}
	\sqrt{1+e^{2\phi}C_\alpha g^{\alpha\beta}C_{\beta}}}
\epsilon^{\alpha\beta \gamma\delta\omega}H_{\gamma\delta\omega})\times \nonumber \\
& &\times (e^{-\frac{4}{3}\phi}+e^{\frac{2}{3}\phi}C_\alpha g^{\alpha\beta}C_\beta) \ . 
\nonumber \\
\end{eqnarray}
%and consequently we obtain that the first square root term in the action has the form 
%\begin{eqnarray}
%e^{-2\phi}\det 
%(g_{\alpha\beta}+\frac{e^{2\phi}C_\alpha c_\beta}{1+e^{2\phi}C_\alpha g^{\alpha\beta}C_\beta}+i\frac{e^{\phi}}{6\sqrt{-\det g}
%	\sqrt{1+e^{2\phi}C_\alpha g^{\alpha\beta}C_\beta}}
%g_{\alpha\alpha'}g_{\beta\beta'}\epsilon^{\alpha'\beta'\gamma\delta\omega}
%H_{\gamma\delta\omega})
%\times \nonumber \\
% (1+e^{2\phi}C_\alpha g^{\alpha\beta}C_\beta) \ . 
%\nonumber \\
%\end{eqnarray}
Let us now analyse the second term in the action (\ref{actM5}) that using (\ref{parta}) takes the form
\begin{eqnarray}
& &-\frac{\sqrt{-\hg}}{4\widehat{\partial a\partial a}}
\hH^{*mn}H_{mnr}\partial^ra
=-\frac{\sqrt{-\hg}}{4\hg^{yy}}\hH^{*\alpha\beta}H_{\alpha\beta\gamma}
\hg^{\gamma 5}=\nonumber \\
%=-\frac{\sqrt{-\hg}}{24\sqrt{-\hg}\hg^{yy}}
%\epsilon^{\alpha\beta\gamma\delta\omega}H_{\gamma\delta\omega}H_{\alpha\beta\gamma}
%\hg^{\gamma y}=
& &=-\frac{e^{\frac{2}{3}\phi}}{24(e^{-\frac{4}{3}\phi}+e^{\frac{2}{3}\phi}
	C_\alpha C^\alpha)}\epsilon^{\alpha\beta\gamma\delta
\omega}H_{\gamma\delta\omega}H_{\alpha\beta\rho}g^{\rho\sigma}C_\sigma \ . 
\nonumber \\
\end{eqnarray}
Collecting these terms together we finally obtain   dimensionally reduced form of the action
\begin{eqnarray}\label{Sdimspec}
& &S=-T_{D4}\int d^5\xi
e^{-\phi}\sqrt{-\det 
	(g_{\alpha\beta}-
	\frac{C_\alpha C_\beta }{
		1+e^{2\phi}C_\gamma g^{\gamma\delta}C_\delta}+
	i\frac{e^{\phi}g_{\alpha\alpha'}g_{\beta\beta'}}{6\sqrt{-\det g}
		\sqrt{1+e^{2\phi}C_\gamma g^{\gamma\delta}C_\delta
}}
	\epsilon^{\alpha'\beta' \gamma\delta\omega}H_{\gamma\delta\omega})}\times \nonumber \\
& &\sqrt{1+e^{2\phi}C_\gamma g^{\gamma\delta}C_\delta}
\nonumber \\
& &-\frac{e^{2\phi}}{24(1+e^{2\phi}
	C_\gamma g^{\gamma\delta} C_\delta)}\epsilon^{\alpha\beta\gamma\delta
	\omega}H_{\gamma\delta\omega}H_{\alpha\beta\rho}g^{\rho\sigma}C_\sigma+
\nonumber \\
& &+\frac{1}{6}\int d^5\xi V^{\alpha\beta\gamma}(H_{\alpha\beta\gamma}-
(\partial_\alpha B_{\beta\gamma}+\partial_\beta B_{\gamma\alpha}+
\partial_\gamma B_{\alpha\beta})) \ , \nonumber\\
\end{eqnarray}
where we identified $T_{D4}\equiv T_{M5}\int dy$ since we anticipate that the action above corresponds to D4-brane action in Type IIA theory. However this correspondence
is not quite correct since 
the action (\ref{Sdimspec}) still depends on the field strength 
of two  form field $B_{\alpha\beta}$ while DBI action is defined using gauge field which is one form. In order to find such an action we should perform duality transformation following \cite{Tseytlin:1996it,Aganagic:1997zk}. The first step is to extend the action as in (\ref{Sdimspec}) that allows us to 
treat  $B_{\beta\gamma}$ and $H_{\alpha\beta\gamma}$ as independent variables. Since now $B_{\alpha\beta}$ is independent it can be integrated out that  implies 
\begin{equation}
\partial_\alpha V^{\alpha\beta\gamma}=0 \ . 
\end{equation}
This equation has solution in five dimensions in the form
\begin{equation}
V^{\alpha\beta\gamma}=
\epsilon^{\alpha\beta\gamma\delta\omega}\partial_\delta A_\omega \ .
\end{equation}
In order to find dual action we should integrate out three form $H_{\alpha\beta\gamma}$. To do this we will perform inverse procedure to the analysis presented in \cite{Tseytlin:1996it,Aganagic:1997zk}. Explicitly,  we start with the action (\ref{Sdimspec}) and find its dual formulation. Following \cite{Tseytlin:1996it,Aganagic:1997zk} we use five dimensional Lorentz invariance and  switch to local inertial frame where $g_{\alpha\beta}=\eta_{\alpha\beta}$ and where we also presume that $C_\alpha$ has non-zero component $C_0$ only. We further presume that there are non-zero components $H_{340}$ and $H_{120}$ so that the matrix 
\begin{equation}
\bM_{\alpha\beta}\equiv 
g_{\alpha\beta}-
\frac{e^{2\phi}C_\alpha C_\beta }{
	1+e^{2\phi}C_\gamma C^\gamma}+
i\frac{e^{\phi}g_{\alpha\alpha'}g_{\beta\beta'}}{6\sqrt{-\det g}
	\sqrt{1+e^{2\phi}C_\gamma C^
		\gamma}}
\epsilon^{\alpha'\beta' \gamma\delta\omega}H_{\gamma\delta\omega}
\end{equation}
has the form 
\begin{equation}
\bM_{\alpha\beta}=\left(\begin{array}{ccccc}
-1-\frac{e^{2\phi}C_0^2}{1-e^{2\phi}C_0^2} & 0 & 0 & 0 & 0 \\
0 & 1 & i\frac{e^{\phi}}{\sqrt{1-e^{2\phi}C_0^2}}H_{340} & 0 & 0 \\
0 &-i\frac{e^{\phi}}{\sqrt{1-e^{2\phi}C_0^2}}H_{340} & 1 & 0 & 0 \\
0 & 0 & 0 & 1 & i\frac{e^{\phi}}{\sqrt{1-e^{2\phi}C_0^2}}H_{120} \\
0 & 0 & 0 & -i\frac{e^{\phi}}{\sqrt{1-e^{2\phi}C_0^2}}H_{120}  & 1 \\
\end{array}\right)
\end{equation}
so that $\det\bM_{\alpha\beta}$ is equal to 
\begin{equation}\det\bM_{\alpha\beta}
=-\left(1+\frac{e^{2\phi}C_0^2}{1-e^{2\phi}C_0^2}\right)
\left(1-\frac{e^{2\phi}}{1-e^{2\phi}C_0^2}H_2^2\right)
\left(1-\frac{e^{2\phi}}{1-e^{2\phi}C_0^2}H_1^2\right) \ , 
\end{equation}
where $H_{340}=H_2 \ , H_{120}=H_1$. We also introduce notation 
$V^{120}=V_1, V^{340}=V_2$. Then the action has the form
\begin{eqnarray}\label{Sred}
%	S=
%	-T_{D5}\int d^5\xi e^{-\phi}
%\sqrt{\left(1+\frac{e^{2\phi}C_0^2}{1-e^{2\phi}C_0^2}\right)
%\left(1-\frac{e^{2\phi}}{1-e^{2\phi}C_0^2}H_2^2\right)
%\left(1-\frac{e^{2\phi}}{1-e^{2\phi}C_0^2}H_1^2\right)}\sqrt{1-e^{2\phi}C_0^2}+\nonumber \\
%+T_{D5}\int d^5\xi (\frac{e^{2\phi}}{1-e^{2\phi}C_0^2}H_1H_2+V_1H_1+V_2H_2)=
%\nonumber \\
& &	S=-T_{D4}\int d^5\xi e^{-\phi}
\sqrt{
	\left(1-\frac{e^{2\phi}}{1-e^{2\phi}C_0^2}H_2^2\right)\left(1-\frac{e^{2\phi}}{1-e^{2\phi}C_0^2}H_1^2\right)}+\nonumber \\
& &+T_{D4}\int d^5\xi \left(\frac{e^{2\phi}C_0}{1-e^{2\phi}C_0^2}H_1H_2+V_1H_1+V_2H_2\right) \ . 
\nonumber \\
\end{eqnarray}
Let us solve equations of motion for $H_1$ and $H_2$ that follow from (\ref{Sred}) 
\begin{eqnarray}\label{eqH}
\triangle e^{-\phi}\sqrt{\frac{1-\triangle H_2^2}{1-\triangle H_1^2}}H_1
	+\triangle C_0 H_2+V_1=0 \ , \nonumber \\
	\triangle e^{-\phi}\sqrt{\frac{1-\triangle H_1^2}{1-\triangle H_2^2}}H_2
		+\triangle C_0 H_1+V_2=0 \ , \nonumber \\
\end{eqnarray}
where 
\begin{equation}
\triangle=\frac{e^{2\phi}}{1-e^{2\phi}C_0^2} \ . 
\end{equation}
The equations (\ref{eqH}) have solutions
%Then solving these equations of motion we obtain 
%\begin{equation}
%\triangle^2 e^{-2\phi}H_1H_2=(\triangle C_0 H_2+V_1)(\triangle C_0 H_1+V_2)
%\Rightarrow H_2=\frac{1}{\triangle (H_1-C_0V_2)}(V_1\triangle C_0 H_1+V_2V_1)
%\end{equation}
%and hence
%\begin{equation}
%\triangle C_0 H_2+V_1=\frac{e^{-2\phi}\triangle H_1V_1}{H_1-C_0 V_2}
%\end{equation}
%Then from the first equation we find
%\begin{eqnarray}
%(\triangle C_0 H_2+V_1)^2=\triangle^2 e^{-2\phi}
%(\frac{1-\triangle H_2^2}{1-\triangle H_1^2})H_1^2
%\nonumber \\
%e^{-2\phi}V_1^2(1-\triangle H_1^2)-(H_1-C_0 V_2)^2=- \triangle H_2^2(H_1-C_0V_2)^2 \Rightarrow 
%\nonumber \\
%\triangle e^{-2\phi}V_1^2(1-\triangle H_1^2)-\triangle H_1^2+2\triangle H_1C_0V_2-\triangle C_0^2V_2^2=
%-\triangle^2 V_1^2C_0^2H_1^2-2V_1^2V_2\triangle C_0 H_1-V_2^2V_1^2 \nonumber \\
%H_1^2\triangle(-\triangle e^{-2\phi}V_1^2-1+\triangle V_1^2 C_0^2)
%+2H_1\triangle C_0 V_2(1+V_1^2)
%+\triangle e^{-2\phi}V_1^2-\triangle C_0^2 V_2^2+V_2^2V_1^2=0\nonumber \\
%\Rightarrow 
%-H_1^2\triangle (1+V_1^2)+2H_1\triangle C_0V_2(1+V_1^2)+
%\nonumber \\
%\end{eqnarray}
%that has solution 
\begin{eqnarray}\label{solH1H2}
%H_1=\frac
%{-\triangle V_2 C_0(1+V_1^2)\pm e^{-\phi}
%\sqrt{\triangle (1+V_1^2)}\sqrt{\triangle (1+V_2^2)}V_1}
%{-\triangle (1+V_1^2)}=\nonumber \\
H_1=V_2C_0\mp e^{-\phi}\sqrt{\frac{1+V_2^2}{1+V_1^2}}V_1 \ , \quad 
H_2=V_1C_0\mp e^{-\phi}\sqrt{\frac{1+V_1^2}{1+V_2^2}}V_2 \ . \nonumber \\
\end{eqnarray}
Finally inserting these solutions into (\ref{Sred}) we obtain
\begin{eqnarray}\label{Sdual}
%S-T_{D5}\int d^5\xi e^{-\phi}
%\sqrt{
%	(1-\triangle H_2^2)(1-\triangle H_1^2)}
%+T_{D5}\int d^5\xi (C_0\triangle H_1H_2+V_1H_1+V_2H_2)=
%\nonumber \\
%=T_{D5}\int d^5\xi (C_0\frac{H_2}{H_1}+\frac{V_1}{\triangle H_1}+V_2H_2)=\nonumber \\
%=T_{D5}\int d^5\xi (\frac{e^{-2\phi}V_1}{H_1-C_0 V_2}+V_2H_2)=\nonumber \\
S=-T_{D4}\int d^5\xi [e^{-\phi}\sqrt{(1+V_1^2)(1+V_2^2)}-C_0 V_1V_2] \nonumber \\
\end{eqnarray}
using that on-shell we have
\begin{equation}
-e^{-\phi}\sqrt{(1-\triangle H_1^2)(1-\triangle H_2^2)}=\frac{1}{\triangle H_1}
(\triangle C_0 H_2+V_1)(1-\triangle H_1^2) \nonumber \\
\end{equation}
and we have chosen $-$ sign in (\ref{solH1H2}) to have an action with correct sign. Then we finally  introduce   $F_{\alpha\beta}=\partial_\alpha A_\beta-\partial_\beta A_\alpha$ and restore world-volume covariance so that  the action (\ref{Sdual}) has final form 
\begin{equation}
S=-T_{D4}\int d^5\xi \left[e^{-\phi}
\sqrt{-\det (g_{\alpha\beta}+F_{\alpha\beta})}-\frac{1}{8}
\epsilon^{\alpha\beta\gamma\delta\omega}C_\alpha F_{\beta\gamma}F_{\delta\omega}\right] \ . 
\end{equation}
This is action for D4-brane in the background with metric $g_{
\mu\nu}$ and Ramond-Ramond form $C_\mu$. Observe that we proceeded in the inverse
direction to the duality transformation studied in \cite{Aganagic:1997zk}
where the starting point was D4-brane in the background with non-trivial metric and Ramond-Ramond one form.  In the next section we apply similar idea to the case of null dimensional reduction of M5-brane.
\section{Null Dimensional Reduction of M5-brane}\label{third}
We begin this section with the review of basic facts about
null dimensional reduction of  gravity.

Let us consider gravity background with  null isometry and perform 
its dimensional reduction, following 
\cite{Julia:1994bs,Bergshoeff:2017dqq}. We  presume that we have vierbein $\hE_M^{ \ A}$
and inverse $\hE^M_{ \ A}$ where $M=0,1,\dots,10$ and tangent space-indices
$A,B=0,1,\dots,10$ that obey
\begin{equation}
\hE_M^{ \ A}\hE^M_{ \ B}=\delta^A_B \ , \quad 
\hE_M^{ \ A}\hE^N_{ \ A}=\delta_M^N \ . 
\end{equation}
If we try to dimensionally reduce along null direction we assume an existence
of null Killing vector $\xi=\xi^M\partial_M$ that obeys
\begin{equation}
\mL_\xi \hg_{MN}=0 \ , \quad  \xi^M \hg_{MN}\xi^N=0 \ . 
\end{equation}
Without loss of generality we choose adapted coordinates $x^M=(x^\mu,y)$ where 
$\mu=0,1,\dots,9$ and where $x^{10}=y$ so that $\xi=\xi^y\partial_y$. Then $\mL_\xi \hg=0$ implies that $\hg$ is independent on $y$ and null condition on $\xi$ implies
\begin{equation}
%\xi^y \hg_{yy}\hg^y=0\Rightarrow 
\hg_{yy}=0 \ . 
\end{equation}
To proceed further we introduce suitable ansatz for vierbein, following \cite{Julia:1994bs,Bergshoeff:2017dqq}. We split $11$ tangent space indices into $a=1,\dots,9$ and remaining as $+,-$ so that $\eta_{ab}=\delta_{ab} \ , \eta_{+-}=\eta_{-+}=1$. Then we have following ansatz for $\hE_M^{ \ A}$
\begin{eqnarray}
& &\hE_\mu^{ \ a}=e_\mu^{ \ a} \ , \quad \hE_\mu^{ \ -}=S^{-1}\tau_\mu \ , \quad 
\hE_\mu^{ \ +}=-Sm_\mu \ , \nonumber \\
& &\hE_y^{ \ a}=0 \  ,\quad \hE_y^{ \ -}=0 \ , \quad \hE_y^{ \ +}=S \ \nonumber \\
\end{eqnarray}
and hence we have following components of metric
\begin{eqnarray}
\hg_{\mu\nu}=
%\hE_\mu^{ \ a}\hE_\nu^{ \ b}\delta_{ab}+\hE_\mu^{ \ +}\hE_\nu^{ \ -}\eta_{+-}+\hE_\mu^{ \ -}\hE_\nu^{ \ +}\eta_{-+}=
h_{\mu\nu}-m_\mu\tau_\nu-\tau_\mu m_\nu \ , \quad 
\hg_{\mu y}=\tau_\mu=\hg_{y\mu} \ , \quad \hg_{yy}=0 \ . \nonumber \\
%\hE_\mu^{ \ a}\hE_v^{ \ b}\delta_{ab}+\hE_\mu^{ \ +}\hE_v^{ \ -}\eta_{+-}
%+\hE_\mu^{ \ -}\hE_\nu^{ \ +}\eta_{-+}=\tau_\mu=\hE_{v\mu}\nonumber \\
%\hg_{vv}=\hE_v^{ \ a}\hE_v^{ \ b}\delta_{ab}+
%\hE_v^{ \ +}\hE_v^{ \ -}\eta_{+-}+
%\hE_v^{ \ -}\hE_v^{ \ +}\eta_{-+}=0
\end{eqnarray}
Let us now consider induced metric on the world-volume of M5-brane with world-volume coordinates $m,n=0,1,\dots,5$ when we presume that M5-brane is extended along $y-$direction. Explicitly
\begin{equation}
\xi^5=y \ 
\end{equation}
so that the  induced metric $\hg_{mn}$ has the form 
\begin{eqnarray}
\hg_{\alpha\beta}=h_{\alpha\beta}-m_\alpha \tau_\beta-\tau_\alpha m_\beta\equiv \bh_{\alpha\beta} \ , \quad  
\hg_{yy}=0 \ , \quad \hg_{y\alpha}=\tau_\alpha \ , \nonumber \\
\end{eqnarray}
where $h_{\alpha\beta}=h_{\mu\nu}\partial_\alpha X^\mu\partial_\beta X^\nu \ , 
\quad m_\alpha=m_\mu\partial_\alpha X^\mu \ , \quad \tau_\alpha=\tau_\mu \partial_\alpha X^\mu$. 
For further purposes we also write elements of the world-volume  inverse metric in the form 
\begin{equation}
\hg^{yy}=2\Phi \ ,  \quad 
\hg^{\alpha y}=-\hv^\alpha \ ,  \quad 
\hg^{y\alpha}=-\hv^\alpha , \quad   \hg^{\alpha\beta}=h^{\alpha\beta} \ , 
\end{equation}
where we defined
\begin{equation}
\hv^\alpha=v^\alpha-h^{\alpha\beta}m_\beta \ , \quad  
\Phi=-v^\alpha m_\alpha+\frac{1}{2}m_\alpha h^{\alpha\beta}m_\beta \ , 
\end{equation}
and where we also introduced $v^\alpha,h^{\alpha\beta}$ through the relations
\begin{equation}
\tau_\alpha v^\alpha=-1 \ , \quad  \tau_\alpha h^{\alpha\beta}=0 \ , \quad  
v^\alpha h_{\alpha\beta}=0 \ ,  \quad h_{\alpha \beta}h^{\beta\gamma}-\tau_\alpha
v^\gamma-=\delta_\alpha^\gamma \ . 
\end{equation}
It is important to stress that $v^\alpha$ and $h^{\alpha\beta}$ are  five dimensional vector and tensor respectively. 

Let us now proceed to the null dimensional reduction of M5-brane action 
(\ref{actM5}). First of all we again use the fact that generally we have
\begin{eqnarray}\label{genbra}
& &-\det\left(\hg_{mn}+i\frac{\hH^*_{mn}}{\sqrt{\widehat{\partial a\partial a}}}\right) \nonumber \\
%=-\det(\hg_{mn}+i\frac{1}{6\sqrt{-\hg}}\hg_{mr}\hg_{ns}
%\epsilon^{rstuvw}H_{tuv}\partial_w a)=\nonumber \\
%=-\det (\hg_{mr}(\hg^{rs}+i\frac{1}{6\sqrt{-\hg}}
%\epsilon^{rstuvw}H_{tuv}\partial_w a)\hg_{sn})=\nonumber \\
& &=-\det \hg_{mr}\left(-\det (\hg^{rs}+i\frac{1}{6\sqrt{\widehat{\partial a\partial a}}\sqrt{-\hg}}
\epsilon^{rstuvw}H_{tuv}\partial_w a)\right)(-\det\hg_{sn}) \ . 
\nonumber \\
\end{eqnarray}
As in previous section we find that non-zero components of $H_{mnr}$ are
$H_{\alpha\beta\gamma}$ when we also imposed the gauge fixing condition 
(\ref{parta}). Using these facts we find that the second bracket in (\ref{genbra}) is 
equal to
%Let us denote remaining world-volume fields as $\alpha,\beta=0,1,2,3,4$. Then in the gauge $B_{5\alpha}=0$ we have non-zero components of $H_{mnr}$ to be equal to
%\begin{equation}
%H_{\alpha\beta\gamma} \ , 
%H_{\alpha\beta 5}=H_{\alpha\beta}
%\end{equation}
%Then by definition
%\begin{equation}
%H_{\alpha\beta}=2(\partial_5 B_{mn}+\partial_\alpha B_{\beta 5}+
%\partial_\beta B_{5m})=0
%\end{equation}
%using the fact that $B_{5m}=0$ and $\partial_5 A_{\alpha\beta}=0$. Since $\partial_w a=\delta_w^5$ we obtain that $\epsilon^{rstuvw}$ has non-zero components
%$\epsilon^{\alpha\beta\gamma\delta\omega 5}=\equiv \epsilon^{\alpha\beta\gamma\delta\omega}$. 
%
%As we argued above $\det\hg_{mn}$ is equal to 
%\begin{equation}
%\det \hg_{mn}=\det(-\tau_\alpha\tau_\beta+h_{\alpha\beta}) \ . 
%\end{equation}
%and also
\begin{eqnarray}\label{exphelp}
& &-\det (\hg^{rs}+i\frac{1}{6\sqrt{\widehat{\partial a\partial a}}\sqrt{-\hg}}
\epsilon^{\alpha\beta\gamma\delta\omega}H_{\gamma\delta\omega})=
-\left|\begin{array}{cc}
h^{\alpha\beta}+i\frac{1}{6\sqrt{2\Phi}\sqrt{-\hg}}
\epsilon^{\alpha\beta\gamma\delta\omega}H_{\gamma\delta\omega} & -\hv^\beta \\
-\hv^\alpha & 2\Phi \\ \end{array}\right|=
\nonumber \\
& &=-2\Phi\det (h^{\alpha\beta}-\frac{1}{2\Phi}\hv^\alpha\hv^\beta
+i\frac{1}{6\sqrt{-\det\bh_{\alpha\beta}}}
\epsilon^{\alpha\beta\gamma\delta\omega}H_{\gamma\delta\omega}) \ , 
\nonumber \\
\end{eqnarray}
where in the final step we used the fact that 
\begin{equation}
\det\hg_{mn}=\left|\begin{array}{cc}
0 & -\tau_\beta \\
-\tau_\alpha & \barh_{\alpha\beta} \\ \end{array}\right|=
\left|\begin{array}{cc}
-\tau_\gamma \barh^{\gamma\delta}\tau_\delta & 0 \\
-\tau_\alpha & \barh_{\alpha\beta} \\ \end{array}\right|=
\frac{1}{2\Phi}\det \barh_{\alpha\beta} \  
\end{equation}
together with the fact that the matrix
\begin{equation}
\bar{h}^{\alpha\beta}=h^{\alpha\beta}-\frac{1}{2\Phi}\hv^\alpha \hv^\beta \  
\end{equation}
is  inverse to $\bar{h}_{\alpha\beta}=h_{\alpha\beta}-m_\alpha
\tau_\beta-\tau_\alpha m_\beta$
\begin{eqnarray}
\bh_{\alpha\beta}\bh^{\beta\gamma}=\delta_\alpha^\gamma \ .
\end{eqnarray}
This can be easily proved using the fact that 
\begin{equation}
\bh_{\alpha\beta}\hv^\beta=2\tau_\alpha \Phi \ , \quad \tau_\alpha \barh^{\alpha\beta}\tau_\beta=-\frac{1}{2\Phi} \ . 
\end{equation}
With the help of these results we obtain that the gauge fixed
form of the  expression (\ref{exphelp}) is equal to
\begin{eqnarray}
%-\det \hg_{mr}(-\det (\hg^{rs}+i\frac{1}{6\sqrt{-\hg}}
%\epsilon^{rstuvw}H_{tuv}\partial_w a))(-\det\hg_{sn})
%\nonumber \\
-\det \left(\hg_{mn}+i\frac{\hH^*_{mn}}{\sqrt{\widehat{\partial a\partial a}}}\right)=-\frac{1}{2\Phi}
\det \left(\bh_{\alpha\beta}+i\frac{1}{6\sqrt{-\det\bh_{\alpha\beta}}}\bh_{\alpha\gamma}
\bh_{\beta\delta}
\epsilon^{\gamma\delta\sigma\omega\rho}H_{\sigma\omega\rho}\right) \ . 
\nonumber \\
\end{eqnarray}
Further we have
\begin{eqnarray}
-\frac{\sqrt{-\hg}}{4\widehat{\partial a\partial a}}
\hH^{*mn}H_{mnr}\partial^r a=
%-\frac{\sqrt{-\hg}}{4 \hg^{55}}
%H^{*\alpha\beta}H_{\alpha\beta\gamma}\hg^{\gamma 5}=
%\nonumber \\
%=\frac{1}{48\Phi}
%\hH^{*\alpha\beta}H_{\alpha\beta\gamma}\hv^\gamma=
%\nonumber \\
%=\frac{1}{48\Phi}
%\epsilon^{\alpha\beta\gamma\delta\omega}H_{\gamma\delta\omega}
%H_{\alpha\beta\sigma}\hv^\sigma= \nonumber \\
\frac{1}{48
\Phi}\epsilon^{\alpha\beta\gamma\delta\omega}H_{\gamma\delta\omega}
H_{\alpha\beta\sigma}\hv^\sigma \nonumber \\
\end{eqnarray}
so that we obtain M5-brane action dimensionally reduced along null direction in the form
\begin{eqnarray}
& &S=-T_{M5}\int dy\int d^5\xi \left[\sqrt{-\frac{1}{2\Phi}\det (\bh_{\alpha\beta}+i\frac{1}{6\sqrt{-\det\bh_{\alpha\beta}}}
\bh_{\alpha\gamma}\bh_{\beta\delta}\epsilon^{\gamma\delta\sigma\omega\rho}
H_{\sigma\omega\rho})} \right.
\nonumber \\					
& & \left.+\frac{1}{48
	\Phi}\epsilon^{\alpha\beta\gamma\delta\omega}H_{\gamma\delta\omega}
H_{\alpha\beta\sigma}\hv^\sigma \right] \ . \nonumber \\
\end{eqnarray}
It is important to stress that $\Phi$ is equal to $-\frac{1}{2}\hv^\alpha\bh_{\alpha\beta}\hv^\beta$. However it is convenient to treat it as independent  when we add  $\int d^5\xi B(2\Phi+\hv^\alpha\bh_{\alpha\beta}\hv^\beta)$ into action. Further, this action still depend on two form $B_{\alpha\beta}$ however it is more natural to find its dual formulation in the same way as in case of the spatial dimensional reduction. In order to find such formulation we 
again extend the action as in previous section
\begin{eqnarray}\label{extactnull}
& &S=-T_{ND4}\int d^5\xi \left[\sqrt{-\frac{1}{2\Phi}\det (\bh_{\alpha\beta}+i\frac{1}{6\sqrt{-\det\bh_{\alpha\beta}}}
	\bh_{\alpha\gamma}\bh_{\beta\delta}\epsilon^{\gamma\delta\sigma\omega\rho}
	H_{\sigma\omega\rho})} \right.
\nonumber \\					
& & \left.+\frac{1}{48
	\Phi}\epsilon^{\alpha\beta\gamma\delta\omega}H_{\gamma\delta\omega}
H_{\alpha\beta\sigma}\hv^\sigma +B(2\Phi+\hv^\alpha\bh_{\alpha\beta}\hv^\beta)\right]
+\nonumber \\
& &+\frac{1}{6}\int d^5\xi V^{\alpha\beta\gamma}(H_{\alpha\beta\gamma}-
(\partial_\alpha B_{\beta\gamma}+\partial_\beta B_{\gamma\alpha}+
\partial_\gamma B_{\alpha\beta})) \ ,  \nonumber\\
\end{eqnarray}
where we introduced effective non-relativistic D4-brane tension through
the relation
\begin{equation}
T_{ND4}=T_{M5}\int dy \ . 
\end{equation}
In (\ref{extactnull}) we 
treat $H_{\alpha\beta\gamma}$ and $B_{\alpha\beta}$ as
independent fields. Then the equations of motion for  $B_{\alpha\beta}$ implies
that $V^{\alpha\beta\gamma}$ is equal to 
\begin{equation}
V^{\alpha\beta\gamma}=
\epsilon^{\alpha\beta\gamma\delta\omega}\partial_\delta A_\omega \ . 
\end{equation}
Finally we should solve equations of motion for $H_{\alpha\beta\gamma}$. Since it
is difficult to solve them in the full generality we proceed in the similar way as in previous section even if we have to be more careful now. First of all we can still
presume that the space metric $h_{\alpha\beta}$ has the form
\begin{equation}
h_{\alpha\beta}=\delta_\alpha^i\delta_\beta^j\delta_{ij} \ , \quad  i,j=1,\dots,4 \ .
\end{equation}
On the other hand in case of time one form we presume that it is constant
and equal to 
\begin{equation}
\tau_\alpha=\tau \delta_\alpha^0 \ . 
\end{equation}
We further  presume that $m$ has non-zero components $m_0,m_1$ so that $\bh_{00}=-
2\tau m_0, \bh_{01}=-\tau m_1=\bh_{10}$.
 Then since $v^\alpha$ is defined as $\tau_\alpha v^\alpha=-1$ 
we have $v^0=-\frac{1}{\tau}$. 
%Note also that  potential $\Phi$ is equal to
%\begin{equation}
%\Phi=\frac{m_0}{\tau}+\frac{1}{2}m_1^2 \ . 
%\end{equation} 
Finally, as in previous section we presume that non-zero components of $H_{\alpha\beta\gamma}$ are $H_{340}\equiv H_2,H_{120}\equiv H_1$. Then the matrix
 $
M_{\alpha\beta}=\bh_{\alpha\beta}+i\frac{1}{6\sqrt{-\det\bh_{\alpha\beta}}}
\bh_{\alpha\gamma}\bh_{\beta\delta}\epsilon^{\gamma\delta\sigma\omega\rho}
H_{\sigma\omega\rho}$ has the form 
\begin{equation}
\left(\begin{array}{ccccc} 
\bh_{00} & \bh_{01} & i\frac{\bh_{01}}{\sqrt{-\bh}}H_2 & 0 & 0 \\
\bh_{10}  & 1 &i\frac{1}{\sqrt{-\bh}}H_{2} & 0 & 0  \\
-i\frac{\bh_{01}}{\sqrt{-\bh}}H_2 & -i\frac{1}{\sqrt{-\bh}}H_{2} & 1 & 0 & 0 \\
0 & 0 & 0 &  1 &  i\frac{1}{\sqrt{-\bh}}H_{1} \\
0 & 0 & 0 & -i\frac{1}{\sqrt{-\bh}}H_{1}  & 1 \\
\\ \end{array}\right)
\end{equation}
where we defined $\bh=\det \bh_{\alpha\beta}=\bh_{00}-\bh_{01}^2$. 
%so that
%\begin{equation}
%\sqrt{-\frac{1}{2\Phi}\det M_{\alpha\beta}}+
%\sqrt{-\frac{\bh_{00}}{2\Phi}(1-\frac{2\Phi}{\bh_{00}} H_2^2)(1+ \frac{2\Phi}{\bh_{00}}H_{1}^2)}
%\end{equation}
Then  the extended action (\ref{extactnull}) has the form 
%\begin{equation}
%S=-T_{ND4}\int d^5\xi [\sqrt{-\frac{\bh}{2\Phi}
%(1+ \frac{1}{\bh}H_2^2)(1+\frac{1}{\bh}H_{1}^2)}+\frac{1}{2\Phi}H_2 H_1v^0-V^1 H_1-V^2H_2
%+B(2\Phi+2\frac{m_0}{\tau}+m_1^2)
%] \ , 
%\end{equation}	
%	where $V_1\equiv V^{120} \ , V_2\equiv V^{340}$. Finally we perform rescaling
%	$H_{1,2}=\sqrt{-\bh}H_{1,2},V_{1,2}=\frac{1}{\sqrt{2\Phi}}\tV_{1,2}$ and hence the action can be written as
\begin{eqnarray}\label{actH12}
& &S=-T_{ND4}\int d^5\xi
\sqrt{-\frac{\bh}{2\Phi}}
\left[\sqrt{(1-\tH_1^2)(1-\tH_2^2)}+\sqrt{-\frac{\bh (v^0)^2}{2\Phi}}\tH_1\tH_2-
\tV_1\tH_1-\tV_2\tH_2\right]+\nonumber \\
& &+T_{ND4}\int d^5\xi B(2\Phi+2\frac{m_0}{\tau}+m_1^2) \ ,  \nonumber \\
\end{eqnarray}
where $V_1\equiv V^{120} \ , V_2\equiv V^{340}$ and where we performed  rescaling
$H_{1,2}=\sqrt{-\bh}\tH_{1,2},V_{1,2}=\frac{1}{\sqrt{2\Phi}}\tV_{1,2}$. 
From (\ref{actH12}) we obtain  following equations of motion
\begin{eqnarray}\label{eqH12}
-\tH_1\sqrt{\frac{1-\tH_2^2}{1-\tH_1^2}}+\triangle \tH_2-\tV_1=0 \ , \nonumber \\
-\tH_2\sqrt{\frac{1-\tH_1^2}{1-\tH_2^2}}+\triangle \tH_1-\tV_2=0 \ , \nonumber \\
\end{eqnarray}
where $\triangle=\sqrt{-\frac{\bh(v^0)^2}{2\Phi}}$.
If we multiply two equations in (\ref{eqH12}) together we obtain
\begin{equation}
\tH_1\tH_2=(\tV_1-\triangle \tH_2)(\tV_2-\triangle \tH_1)
\end{equation}
that allows us to express $\tH_2$ as
\begin{equation}\label{tH2}
\tH_2=\frac{\tV_1\tV_2-\triangle \tH_1\tV_1}{(1-\triangle^2)\tH_1+\triangle \tV_2} \ . 
\end{equation}
Inserting this result into the first equation in (\ref{eqH12}) we obtain 
quadratic equation for $\tH_1$ that has solution
%\begin{equation}
%\tV_1-\triangle \tH_2=
%\frac{\tH_1\tV_1}{(1-\triangle^2)\tH_1+\triangle \tV_2}
%\end{equation}
%Inserting this to the first equation we finally obtain equation
%\begin{eqnarray}
%\tH_1^2(1-\triangle^2)((1-\triangle^2)+\tV_1^2)
%+2\tH_1\triangle [(1-\triangle^2)+ \tV_1^2]\tV_2+\triangle^2\tV_2^2-\tV_1^2\tV_2^2-\tV_1^2=0
%\nonumber \\
%\end{eqnarray}
%so that
\begin{eqnarray}\label{tH1sol}
\tH_1
%=\frac{-\triangle[(1-\triangle^2)+\tV_1^2]\tV_2\pm
%\sqrt{(1-\triangle^2)+\tV_1^2}\sqrt{(1-\triangle^2)+\tV_2^2}\tV_1}
%{(1-\triangle^2)((1-\triangle^2)+\tV_1^2)}=
%\nonumber \\
=\frac{1}{1-\triangle^2}\left[-\triangle \tV_2
\pm \tV_1\sqrt{\frac{(1-\triangle^2)+\tV_2^2}{(1-\triangle^2)+\tV_1^2}}\right] \ . 
\nonumber \\
\end{eqnarray}
Then inserting this result into (\ref{tH2}) we also find $\tH_2$ to be equal to
\begin{equation}\label{tH2sol}
\tH_2=\frac{1}{1-\triangle^2}\left[-\triangle \tV_1\pm \tV_2
\sqrt{\frac{(1-\triangle^2)+\tV_1^2}{(1-\triangle^2)+\tV_2^2}}\right] \ . 
\end{equation}
Finally inserting (\ref{tH1sol}),(\ref{tH2sol}) into the action (\ref{actH12}) 
we  obtain 
%\begin{eqnarray}
%\sqrt{(1-\tH_1^2)(1-\tH_2^2)}+\triangle \tH_1\tH_2-\tV_1\tH_1-\tV_2\tH_2=
%\nonumber \\
%=\frac{(\triangle \tH_2-\tV_1)}{\tH_1}(1-\tH_1^2+\tH_1^2)
%-\tV_2\tH_2=\nonumber \\
%=-\frac{\tV_1}{(1-\triangle^2)\tH_1+\triangle \tV_2}-\tV_2\tH_2=\nonumber \\
%=\mp \sqrt{\frac{(1-\triangle^2)+\tV_1^2}{(1-\triangle^2)+\tV_2^2}}
%+\frac{\triangle}{1-\triangle^2}\tV_1\tV_2\mp
%\frac{1}{1-\triangle^2}\tV_2^2\sqrt{\frac{(1-\triangle^2)+\tV_1^2}{(1-\triangle^2)+\tV_2^2}}=\nonumber \\
%=\mp\frac{1}{1-\triangle^2}\sqrt{((1-\triangle^2)+\tV_1^2)((1-\triangle^2)+\tV_2^2)}
%+\frac{\triangle}{1-\triangle^2}\tV_1\tV_2 \nonumber \\
%\end{eqnarray}
%Now we choose $+$ sign in order to have an action with right sign and hence
\begin{equation}
S=
%T_{ND4}\int d^5\xi [
%\sqrt{-\frac{\bh}{2\Phi}(1+\frac{1}{1-\triangle^2}\tV_1^2)(
%1+\frac{1}{1-\triangle^2}\tV_2^2)}-\frac{1}{1-\triangle^2}\frac{\bh v^0}{2\Phi}\tV_1\tV_2]=
%\nonumber \\
-T_{ND4}\int d^5\xi\left[\sqrt{-\frac{1}{2\Phi}
	(\bh+\bh\hV_1^2)(1+\hV_2^2)}-\frac{1}{2\Phi}\bh v^0
\hV_1\hV_2+
 B(2\Phi+2\frac{m_0}{\tau}+m_1^2)\right] \ ,   
\end{equation}
where we  performed rescalling
$\tV_{1.2}=\sqrt{1-\triangle^2}\hV_{1,2}$
To proceed further let us observe that we can write
%\begin{equation}
%\left|
%\begin{array}{ccc}
%\bh_{00} & \bh_{01} & \bh_{01}V_1 \\
%\bh_{10} & 1 & V_1 \\
%-\bh_{01}V_1 & -V_1 & 1 \\ \end{array}\right|=\bh+\bh V_1^2 \ .
%\end{equation}
%On the other hand using basic properties of determinant we fined that it is equal to
\begin{equation}
%\left|
%\begin{array}{ccc}
%g_{00} & g_{01} & g_{01}V_1 \\
%g_{10} & 1 & V_1 \\
%-g_{01}V_1 & -V_1 & 1 \\ \end{array}\right|=
%\left|
%\begin{array}{ccc}
%g_{00}-g_{10}g_{01} & 0 & 0 \\
%g_{10} & 1 & V_1 \\
%-g_{01}V_1 & -V_1 & 1 \\ \end{array}\right|=
\bh(1+\hV_1^2)=\left|
\begin{array}{ccc}
\bh_{00}-\bh_{01}\bh_{10} & 0 & 0 \\
0 & 1 & \hV_1 \\
0 & -\hV_1 & 1 \\ \end{array}\right|
\end{equation}
so that it is natural to introduce redefined form of the metric  $\bh'_{00}=\bh_{00}-\bh_{01}\bh_{10}$.
Further, we have $\bh v^0=m_0+m_1m_0\equiv m_0'$ introducing redefined $m_0'$. Using these results
we can restore general form of the action for D4-brane in Newton-Cartan background
in the form
%\begin{equation}
%\bh (1+\tV_1^2)=\left|\begin{array}{ccc}
%\bh'_{00} & 0 & 0 \\
%0 & 1 & \tV_1 \\
%0 & -\tV_1 & 1 \\ \end{array}\right|
%\end{equation}
%so that the action has the form
%\begin{equation}
%S=-T_{ND4}\int d^5\xi[\sqrt{-\frac{1}{2\Phi}
%	\bh'_{00}(1+\hV_1^2)(1+\hV_2^2)}-\frac{1}{2\Phi}\bh'_{00} v^0
%\hV_1\hV_2 \ . \nonumber \\
%\end{equation}
%Finally we use
% $\bh'_{00}v^0=
%m_0+m_1m_0\tau\equiv m'_0$ where we defined  $m'_0$. Then we can write
%the action in the covariant form
\begin{equation}\label{D4final}
S=-T_{ND4}\int d^5\xi \left[\sqrt{-\frac{1}{2\Phi}(\bh_{\alpha\beta}+F_{\alpha\beta})}
-\frac{1}{16\Phi}\epsilon^{\alpha\beta\gamma\delta\omega}m_\alpha F_{\beta\gamma}F_{\delta\omega}
+B(2\Phi+\hv^{\alpha}\bh_{\alpha\beta}\hv^\beta)\right] \ . 
\end{equation}
Finally we can integrate $B$ to replace $2\Phi$ with $-\hv^\alpha \bh_{\alpha\beta}\hv^\beta$. Then we can rewrite the action into alternative form 
with the help of following manipulation
\begin{eqnarray}
& &\frac{1}{2\Phi}\det (\bh_{\alpha\beta}+F_{\alpha\beta})=
\frac{1}{2\Phi}\det\bh_{\alpha\gamma}\det(\delta^\gamma_\beta+ \bh^{\gamma\omega}F_{\omega\beta})=\nonumber \\
& &\det (H_{\alpha\gamma})\det (\delta^\gamma_\beta+ \bh^{\gamma\omega}F_{\omega\beta})
=\det (H_{\alpha\beta}+H_{\alpha\gamma}\bh^{\gamma\omega} F_{\omega\beta})=
\nonumber \\
& &=\det (H_{\alpha\beta}+\tF_{\alpha\beta}) \ ,  \nonumber \\
\end{eqnarray}
where we introduced $H_{\alpha\beta}$ and its inverse $H^{\alpha\beta}$ as
\begin{equation}
H_{\alpha\beta}=-\tau_\alpha\tau_\beta+h_{\alpha\beta} \ , \quad 
H^{\alpha\beta}=h^{\alpha\beta}-v^\alpha v^\beta 
\end{equation}
and also $\tF_{\alpha\beta}$ through the relation
\begin{equation}
\tF_{\alpha\beta}=H_{\alpha\gamma}\bh^{\gamma\delta}F_{\delta\beta} \ . 
\end{equation}
Finally we used the fact that 
\begin{eqnarray}
& &\frac{1}{2\Phi}\det \bh_{\alpha\beta}=
\det \left(\begin{array}{cc} 0 &-\tau_\beta \\
-\tau_\alpha &\bh_{\alpha\beta} \\ \end{array}\right)
=
\det \left(\begin{array}{cc} 0 &-\tau_\beta \\
-\tau_\alpha &-\tau_\alpha\tau_\beta+h_{\alpha\beta} \\ \end{array}\right)
\nonumber \\
& &=\det \left(\begin{array}{cc} -\tau_\gamma H^{\gamma\delta}\tau_\delta &0 \\
-\tau_\alpha &H_{\gamma\delta} \\ \end{array}\right)=
\det(H_{\alpha\beta}) \ . \nonumber \\
\end{eqnarray}
Then we obtain that the action (\ref{D4final}) can be rewritten into the form
\begin{eqnarray}\label{D4finalA}
& &S=-T_{ND4}\int d^5\xi \left[\sqrt{-(H_{\alpha\beta}+\tF_{\alpha\beta})}
-\frac{1}{16\Phi}\epsilon^{\alpha\beta\gamma\delta\omega}m_\alpha
\tF_{\beta\gamma}\tF_{\delta\omega} \right. \nonumber \\
& & \left.-\frac{1}{8\Phi}\epsilon^{\alpha\beta\gamma\delta\omega}m_\alpha \tF_{\beta\gamma}\tau_\delta (-v^\sigma \tF_{\sigma\omega}+\frac{1}{2\Phi}(1+v^\delta m_\delta)\hv^\sigma \tF_{\sigma\omega})\right] \ , \nonumber \\ 
%
%-\frac{1}{16\Phi}\epsilon^{\alpha\beta\gamma\delta\omega}m_\alpha H_{\beta\beta'}\bh^{\beta'\rho}F_{\rho \gamma}
%H_{\delta\delta'}\bh^{\delta'\omega'}F_{\omega'\omega}] \nonumber \\
\end{eqnarray}
The action (\ref{D4final}) or its alternative form (\ref{D4finalA}) 
is final form of non-relativistic D4-brane action in the background that arises by null dimensional reduction of M-theory. The kinetic part of the action has standard
DBI like form and has functionally the same form as  D2-brane action in torsional Newton-Cartan background
that was found in \cite{Kluson:2021pux}. There is also an additional WZ like term that has an exceptional property that it depends on the world-volume fields $v^\alpha,h^{\alpha\beta}$.
\section{M5-brane Transverse to Null Direction}\label{fourth}
In this last section we briefly discuss situation when we presume that  M5-brane  is transverse to null direction. 
%
%Let us now consider M5-brane that is transverse to null direction. 
%We again start with the action
%Let us consider M5-brane action in the form 
%\footnote{We work with the metric of the signature 
%	$(-,+,\dots,+)$.}
%\begin{equation}
%S=-\int d^6\xi [\sqrt{-\det (\hg_{mn}+i\frac{\hH^*_{mn}}{\sqrt{\widehat{\partial a\partial a}}})}
%-\frac{\sqrt{-\hg}}{4\widehat{\partial a\partial a}}
%\hH^{*mn}H_{mnr}\partial^ra] \ , 
%\end{equation}
%where
%\begin{equation}
%\widehat{\partial a\partial a}=\partial_m a \hg^{mn}
%\partial_n a
%\end{equation}
%and 
%where
%\begin{equation}
%m,n=0,1,\dots,5 \ , 
%\end{equation}
%are vector indices of $d=6$ world-volume coordinates $\xi^m$,
%\begin{equation}
%M.N=0,1,\dots,10
%\end{equation}
%are vector indices of $D=11$ target space-time coordinates
%$X^M$ 
%\begin{equation}
%\hg_{mn}=\partial_m X^M\hg_{MN}\partial_n X^N
%\end{equation}
%is the world-volume metric that is induced by embedding the five-brane
%into $D=11$ background with metric $\hg_{MN}$
%Further, we have
%\begin{equation}
%\hH^{*}_{mn}=\frac{1}{6\sqrt{-\hg}}\hg_{mr}\hg_{ns}
%\epsilon^{rstuvw}H_{tuv}\partial_r a \ , 
%\end{equation}	
In this case the induced metric has the form
\begin{equation}\label{hgNS5}
\hg_{mn}=\bh_{mn}+\tau_m \partial_n y+\partial_m y \tau_n \ , \quad \tau_m=
\tau_\mu\partial_m X^\mu \ , \quad \bh_{mn}=\bh_{\mu\nu}\partial_m X^\mu\partial_n X^\nu \ ,
\end{equation}
where 
\begin{equation}
\bh_{mn}=h_{mn}-\tau_m m_n-
\tau_n m_m \ . 
\end{equation}
Note that this metric is non-singular with inverse metric
\begin{equation}
\bh^{mn}=h^{mn}-\frac{1}{2\Phi}\hv^m\hv^n \ , 
\end{equation}
where again
\begin{equation}
\hv^m=v^m-h^{mn}m_n \ , \quad \Phi=-v^m m_m+\frac{1}{2}m_m h^{mn}m_n \ , 
\end{equation}
where $v^m$ is six dimensional vector that obeys $v^m\tau_m=-1 \ , v^m h_{mn}=0$. Finally $h^{mn}$ is six dimensional singular matrix that obeys
\begin{equation}
h_{mn}h^{nk}-\tau_m v^k=\delta_m^k \ , \tau_m h^{mn}=0 \ . 
\end{equation}
In order to find NS5-brane in torsional Newton-Cartan background it is appropriate
to express all components of world-volume metric $\hg_{mn}$ and its inverse
$\hg^{mn}$ with the help of $\bh_{mn},\tau_m,m_n$, derivatives of $y$ together with 
inverse fields $\bh^{mn},\hv^m$. In case of $\hg_{mn}$ these relations are given in 
(\ref{hgNS5}). In case of $\hg^{mn}$ the situation is more complicated and after some calculations we find 
\begin{eqnarray}\label{hgNS5in}
%\delta^k_n+\bh^{kp}\bB_{IJ}\bx^I_p \bx^J_n
%=\delta^k_n+\bh^{kp}(\tau_p \bB_{11}\tau_n+\partial_py \bB_{21}\tau_n+
%\tau_p \bB_{12}\partial_n y+\partial_p y \bB_{22}\partial_ny)\nonumber \\
%\hg^{kl}=\bh^{kl}+\bh^{kp}(\tau_p \bB_{11}\tau_n+\partial_py \bB_{21}\tau_n+
%\tau_p \bB_{12}\partial_n y+\partial_p y \bB_{22}\partial_ny)\bh^{nl}=\nonumber \\
& &\hg^{kl}=\bh^{kl}+\frac{1}{1+\frac{1}{\Phi}\hv^m\partial_my+\frac{1}{2\Phi}
	\partial_my h^{mn}\partial_n y}(-\frac{1}{4\Phi^2}\hv^k\hv^l
\partial_m y\bh^{mn}\partial_ny\nonumber \\
& &+\frac{1}{2\Phi}\bh^{kp}\partial_p y(1+\frac{1}{2\Phi}\hv^m\partial_my)\hv^l+
\frac{1}{2\Phi}\hv^k (1+\frac{1}{2\Phi}\hv^m\partial_my)\partial_p y \bh^{pl}-
\frac{1}{2\Phi}\bh^{km}\partial_m y\partial_ny \bh^{nl})  \ .
\nonumber \\
\end{eqnarray}
Then we obtain  an action for non-relativistic NS5-brane when we insert 
(\ref{hgNS5}) and (\ref{hgNS5in})  into  (\ref{actM5}).
 Note that there are two scalar fields $y(\xi),a(\xi)$ that propagate on the world-volume of NS5-brane. In fact this result is similar to transverse dimensional reduction of M5-brane as was studied in 
\cite{Bandos:2000az}. However due to the form of the metric $\hg_{mn},\hg^{mn}$ 
it is impossible to simplify M5-brane action further so that we will not write explicit form of the action for NS5-brane in torsional Newton-Cartan background here. 

%Then $\hg^{mn}$ has the form 
%\begin{equation}
%\hg^{mn}=
%\end{equation}
%using the fact that the metric $\bh_{mn}$ has an inverse metric $\bh^{mn}$ in the form 
%\begin{equation}
%\bh^{mn}=h^{mn}-\frac{1}{2\Phi}\hv^m\hv^n \ , 
%\hv^m=v^m-h^{mn}m_n \ , m_n=m_\mu\partial_m x^\mu \ . 
%\end{equation}

\acknowledgments{This work 
	is supported by the grant “Integrable Deformations”
	(GA20-04800S) from the Czech Science Foundation
	(GACR). }

%%%%%%%%%%%%%%%%%%%%%%%%%%%%%%%%%%%%%%
%%%%%%% Thebibligraphy %%%%%%%%%%%
%%%%%%%%%%
%%%%%%%%%%%%%%%%%%%%%%%%%%%%%%%%%%%%%

\end{document}